\documentclass[prc,aps,twocolumn,showpacs]{revtex4}
\usepackage{amsmath,amssymb}
\usepackage{graphicx}
\def\pcomma{$^,$}
\def\pucla{$^1$}
\def\pmainz{$^2$}
\def\pgwu{$^3$}
\def\pglsg{$^4$}
\def\pkent{$^5$}
\def\pbonn{$^6$}
\def\pgatch{$^7$}
\def\pgiess{$^8$}
\def\ppavia{$^9$}
\def\plpi{$^{10}$}
\def\pdalh{$^{11}$}
\def\psaint{$^{12}$}
\def\pjlab{$^{13}$}
\def\pbasel{$^{14}$}
\def\ptomsk{$^{15}$}
\def\pedinb{$^{16}$}
\def\psackv{$^{17}$}
\def\pinr{$^{18}$}
\def\pzagreb{$^{19}$}
\begin{document}

\title{
A new measurement of the rare decay $\eta \to \pi^0\gamma\gamma$ with
 the Crystal Ball/TAPS detectors at the Mainz Microtron
}
\author{
B.~M.~K.~Nefkens\pucla\footnote[1]{deceased},
S.~Prakhov\pucla\pcomma\pmainz\pcomma\pgwu\footnote[2]{corresponding author, e-mail: prakhov@ucla.edu},
P.~Aguar-Bartolom\'e\pmainz,
J.~R.~M.~Annand\pglsg,
H.~J.~Arends\pmainz,
K.~Bantawa\pkent,
R.~Beck\pbonn,
V.~Bekrenev\pgatch,
H.~Bergh\"auser\pgiess,
A.~Braghieri\ppavia,
W.~J.~Briscoe\pgwu,
J.~Brudvik\pucla,
S.~Cherepnya\plpi,
R.~F.~B.~Codling\pglsg,
C.~Collicott\pdalh\pcomma\psaint,
S.~Costanza\ppavia,
I.~V.~Danilkin\pjlab,
A.~Denig\pmainz,
B.~Demissie\pgwu,
M.~Dieterle\pbasel,
E.~J.~Downie\pmainz\pcomma\pgwu,
P.~Drexler\pgiess,
L.~V.~Fil'kov\plpi,
A.~Fix\ptomsk,
S.~Garni\pbasel,
D.~I.~Glazier\pglsg\pcomma\pedinb,
R.~Gregor\pgiess,
D.~Hamilton\pglsg,
E.~Heid\pmainz\pcomma\pgwu,
D.~Hornidge\psackv,
D.~Howdle\pglsg,
O.~Jahn\pmainz,
T.~C.~Jude\pedinb,
V.~L.~Kashevarov\pmainz\pcomma\plpi,
A.~K\"aser\pbasel,
I.~Keshelashvili\pbasel,
R.~Kondratiev\pinr,
M.~Korolija\pzagreb,
M.~Kotulla\pgiess,
A.~Koulbardis\pgatch,
S.~Kruglov\pgatch\footnotemark[1],
B.~Krusche\pbasel,
V.~Lisin\pinr,
K.~Livingston\pglsg,
I.~J.~D.~MacGregor\pglsg,
Y.~Maghrbi\pbasel,
J.~Mancel\pglsg,
D.~M.~Manley\pkent,
E.~F.~McNicoll\pglsg,
D.~Mekterovic\pzagreb,
V.~Metag\pgiess,
A.~Mushkarenkov\ppavia,
A.~Nikolaev\pbonn,
R.~Novotny\pgiess,
M.~Oberle\pbasel,
H.~Ortega\pmainz,
M.~Ostrick\pmainz,
P.~Ott\pmainz,
P.~B.~Otte\pmainz,
B.~Oussena\pmainz,
P.~Pedroni\ppavia,
A.~Polonski\pinr,
J.~Robinson\pglsg,
G.~Rosner\pglsg,
T.~Rostomyan\pbasel,
S.~Schumann\pmainz,
M.~H.~Sikora\pedinb,
A.~Starostin\pucla,
I.~I.~Strakovsky\pgwu,
T.~Strub\pbasel,
I.~M.~Suarez\pucla,
I.~Supek\pzagreb,
C.~M.~Tarbert\pedinb,
M.~Thiel\pgiess,
A.~Thomas\pmainz,
M.~Unverzagt\pmainz\pcomma\pbonn,
D.~P.~Watts\pedinb,
D.~Werthm\"uller\pbasel,
and L.~Witthauer\pbasel
\\
\vspace*{0.1in}
(A2 Collaboration at MAMI)
\vspace*{0.1in}
}
\affiliation{
\pucla University of California Los Angeles, Los Angeles, California 90095-1547, USA}
\affiliation{
\pmainz Institut f\"ur Kernphysik, University of Mainz,
D-55099 Mainz, Germany}
\affiliation{
\pgwu The George Washington University, Washington, DC 20052-0001, USA}
\affiliation{
\pglsg SUPA, School of Physics and Astronomy, University of Glasgow, Glasgow G12 8QQ, 
United Kingdom}
\affiliation{
\pkent Kent State University, Kent, Ohio 44242-0001, USA}
\affiliation{
\pbonn Helmholtz-Institut f\"ur Strahlen- und Kernphysik, University of Bonn, 
D-53115 Bonn, Germany}
\affiliation{
\pgatch Petersburg Nuclear Physics Institute, 188350 Gatchina, Russia}
\affiliation{
\pgiess II Physikalisches Institut, University of Giessen, D-35392 Giessen, Germany}
\affiliation{
\ppavia INFN Sesione di Pavia, I-27100 Pavia, Italy}
\affiliation{
\plpi Lebedev Physical Institute, 119991 Moscow, Russia}
\affiliation{
\pdalh Dalhousie University, Halifax, Nova Scotia B3H 4R2, Canada}
\affiliation{
\psaint Saint Mary's University, Halifax, Nova Scotia B3H 3C3, Canada}
\affiliation{
\pjlab Jefferson Laboratory, Newport News, Virginia 23606, USA}
\affiliation{
\pbasel Department of Physics, University of Basel, CH-4056 Basel, Switzerland}
\affiliation{
\ptomsk Laboratory of Mathematical Physics, Tomsk Polytechnic University, 634050 Tomsk, Russia}
\affiliation{
\pedinb SUPA, School of Physics, University of Edinburgh, Edinburgh EH9 3JZ, United Kingdom}
\affiliation{
\psackv Mount Allison University, Sackville, New Brunswick E4L 1E6, Canada}
\affiliation{
\pinr Institute for Nuclear Research, 125047 Moscow, Russia}
\affiliation{
\pzagreb Rudjer Boskovic Institute, HR-10000 Zagreb, Croatia}

\date{\today}
                  
\begin{abstract}
 A new measurement of the rare, doubly radiative decay
 $\eta\to\pi^0\gamma\gamma$ was conducted
 with the Crystal Ball and TAPS multiphoton spectrometers 
 together with the photon tagging facility at the Mainz Microtron MAMI.
 New data on the dependence of the partial decay width,
 $\Gamma(\eta\to\pi^0\gamma\gamma)$,
 on the two-photon invariant mass squared, $m^2(\gamma\gamma)$,
 as well as a new, more precise value for the decay width,
 $\Gamma(\eta \to \pi^0\gamma\gamma) = (0.33\pm 0.03_{\mathrm{tot}})$~eV,
 are based on analysis of $1.2\times 10^3 ~\eta \to \pi^0\gamma\gamma$
 decays from a total of $6 \times 10^7$ $\eta$ mesons produced in
 the $\gamma p\to \eta p$ reaction.
 The present results for $d\Gamma(\eta\to\pi^0\gamma\gamma)/dm^2(\gamma\gamma)$
 are in good agreement with previous measurements and recent theoretical calculations
 for this dependence. 
\end{abstract}


\maketitle

\section{Introduction}
 The rare, doubly radiative decay $\eta \to \pi^0\gamma\gamma$
 attracts much interest due to large uncertainties in the experimental
 data and in the calculations based on Chiral Perturbation
 Theory ($\chi$PTh). In many cases, $\chi$PTh
 provides a good description of the low-energy dynamics
 of light pseudoscalar mesons using an expansion in powers
 of small meson masses and momenta~\cite{Gasser}.
 The uncertainties in $\chi$PTh calculations
 for the $\eta\to\pi^0\gamma\gamma$ transition are
 related to the fact that the tree-level amplitudes of
 the lowest orders ${\cal O}({\sf p}^2)$
 (order two in particle four-momentum or masses) and
 ${\cal O}({\sf p}^4)$ vanish.
 The first non-zero contribution comes at ${\cal O}({\sf p}^4)$ either
 from loops involving kaons, strongly suppressed owing to large kaon
 masses, or from pion loops, suppressed as they violate G parity.
 The major contribution to the $\eta\to\pi^0\gamma\gamma$ decay amplitude
 comes from the ${\cal O}({\sf p}^6)$ counterterms that are needed in
 $\chi$PTh to cancel various divergences. The coefficients
 of these counterterms are not determined by $\chi$PTh itself; they
 have to be either adjusted by fitting a model to experimental data
 or estimated using model assumptions,
 e.g. vector-meson dominance (VMD) or Nambu-Jona-Lasinio models~\cite{Th_Models}.
 Since light vector mesons play a critical role in these models,
 the dynamical role of the vector mesons has to be included
 systematically~\cite{Oset_2003,Oset_2008,Ch_Lag_2012} to reach
 a deeper understanding of the $\eta\to\pi^0\gamma\gamma$ decay as well
 as photon-fusion reactions into pseudoscalar mesons
 (e.g., $\gamma\gamma\to\pi^0\pi^0$ and $\gamma\gamma\to\pi^0\eta$).

 To test the models based on $\chi$PTh or to provide an input
 for adjusting their parameters, 
 one has to measure both the decay rate for $\eta\to\pi^0\gamma\gamma$
 and its Dalitz plot, the features of which reflect the decay amplitude.  
 The experimental challenges in measuring
 $\eta\to\pi^0\gamma\gamma\to 4\gamma$ are formidable because of
 the smallness of its decay rate in conjuction with large
 background contributions. This makes the direct measurement of
 the $\eta\to\pi^0\gamma\gamma$ Dalitz plot very difficult;
 it is easier to measure
 the $d\Gamma(\eta\to\pi^0\gamma\gamma)/dm^2(\gamma\gamma)$
 distribution instead.

 The first results on measuring
 the $d\Gamma(\eta\to\pi^0\gamma\gamma)/dm^2(\gamma\gamma)$ dependence, 
 obtained by the Crystal Ball (CB) collaboration at the AGS and by the A2
 collaboration at MAMI-B, were presented in Refs.~\cite{Prakhov_MENU07,CB_AGS}.
 The history of experimental and theoretical efforts
 to measure and calculate the $\eta\to\pi^0\gamma\gamma$ decay width
 is reviewed in Ref.~\cite{CB_AGS} and references therein.
 Another recent attempt~\cite{Lalwani_PhD} to measure the
 $d\Gamma(\eta\to\pi^0\gamma\gamma)/dm^2(\gamma\gamma)$ dependence
 has not been finalized so far.

 The publication of the first experimental data~\cite{Prakhov_MENU07,CB_AGS}
 on $d\Gamma(\eta\to\pi^0\gamma\gamma)/dm^2(\gamma\gamma)$
 inspired the revision of the previous theoretical calculations
 and the adjustment of model parameters according to these data.
 Describing the magnitude of $\Gamma(\eta\to\pi^0\gamma\gamma)$,
 those calculations were not constrained by details of
 the $d\Gamma(\eta\to\pi^0\gamma\gamma)/dm^2(\gamma\gamma)$ shape.
 This situation changed after obtaining the new
 data~\cite{Prakhov_MENU07,CB_AGS}.

 A chiral unitary approach~\cite{Oset_2003} used to calculate
 the $\eta\to\pi^0\gamma\gamma$ decay was revised in Ref.~\cite{Oset_2008}. 
 According to the authors, the largest changes in the calculated decay
 properties were due to using the latest results for 
 radiative widths of vector mesons as input for their model.
 The agreement between the revised calculation of Ref.~\cite{Oset_2008}
 and the first data points~\cite{Prakhov_MENU07} was reasonably good
 for both the shape and the magnitude of
 the $d\Gamma(\eta\to\pi^0\gamma\gamma)/dm^2(\gamma\gamma)$ dependence.
 The integrated partial decay width,
 $\Gamma(\eta\to\pi^0\gamma\gamma)=(0.33\pm0.08)$~eV,
 from the calculation of Ref.~\cite{Oset_2008} is in good agreement
 with the present value, $\Gamma(\eta\to\pi^0\gamma\gamma)=(0.35\pm0.07)$~eV,
 from the Particle Data Group (PDG)~\cite{PDG_2012}.

 A theoretical study of photon-fusion reactions based on a chiral Lagrangian
 with dynamical light vector mesons was recently presented in Ref.~\cite{Ch_Lag_2012}.
 In that work, the chiral Lagrangian contains five unknown constants
 that are relevant for the photon-fusion reactions and parameterize
 the strength of interaction terms involving two vector meson fields.
 These parameters were fitted to the data of three well-known photon-fusion
 reactions and to existing $d\Gamma(\eta\to\pi^0\gamma\gamma)/dm^2(\gamma\gamma)$
 data~\cite{CB_AGS,Prakhov_MENU07,Lalwani_PhD}. The results of that study
 show good agreement with existing data for photon-fusion reactions
 and predict cross sections for poorer known photon-fusion processes.

 This article presents a new measurement of
 the $d\Gamma(\eta\to\pi^0\gamma\gamma)/dm^2(\gamma\gamma)$ dependence
 by the A2 collaboration at MAMI. The results obtained are based
 on the analysis of $1.2\times 10^3 ~\eta \to \pi^0\gamma\gamma$ 
 decays from a total of $6 \times 10^7$ $\eta$ mesons produced
 in the $\gamma p\to \eta p$ reaction,
 allowing the experimental uncertainties in
 measuring the $\eta \to \pi^0\gamma\gamma$ decay to be decreased.
 The $d\Gamma(\eta\to\pi^0\gamma\gamma)/dm^2(\gamma\gamma)$ results
 presented in this work were also used to repeat the analysis of
 photon-fusion reactions~\cite{Ch_Lag_2012,Ch_Lag_refit}.

\section{Experimental setup}
\label{sec:Setup}

The process $\gamma p\to \eta p \to \pi^0\gamma\gamma p$
was measured using the Crystal Ball (CB)~\cite{CB}
as the central spectrometer and TAPS~\cite{TAPS,TAPS2}
as a forward spectrometer. These detectors were
installed in the energy-tagged bremsstrahlung photon beam of
the Mainz Microtron (MAMI)~\cite{MAMI,MAMIC}. 
The photon energies were determined using
the Glasgow-Mainz tagging spectrometer~\cite{TAGGER,TAGGER1,TAGGER2}.

The CB detector is a sphere consisting of 672
optically isolated NaI(Tl) crystals, shaped as
truncated triangular pyramids, which point toward
the center of the sphere. The crystals are arranged in two
hemispheres that cover 93\% of $4\pi$, sitting
outside a central spherical cavity with a radius of
25~cm, which holds the target and inner
detectors. In this experiment, TAPS was
arranged in a plane consisting of 384 BaF$_2$
counters of hexagonal cross section.
It was installed 1.5~m downstream of the CB center
and covered the full azimuthal range for polar angles
from $1^\circ$ to $20^\circ$.
More details on the energy and angular resolution of the CB and TAPS
are given in Refs.~\cite{slopemamic,etamamic}.

 The present measurement used 1508-MeV and 1557-MeV
 electron beams from the Mainz Microtron, MAMI-C~\cite{MAMIC}.
 The data with the 1508-MeV beam were taken in 2007 and those with the 1557-MeV beam
 in 2009. Bremsstrahlung photons, produced by the 1508-MeV electrons
 in a 10-$\mu$m Cu radiator and collimated by a 4-mm-diameter Pb collimator,
 were incident on a 5-cm-long liquid hydrogen (LH$_2$) target located
 in the center of the CB. The energies of the incident
 photons were analyzed up to 1402~MeV by detecting the post-bremsstrahlung
 electrons in the Glasgow-Mainz tagger~\cite{TAGGER,TAGGER1,TAGGER2}.
 With the 1557-MeV electron beam, the incident photons were analyzed up to 1448~MeV,
 and a 10-cm-long LH$_2$ target was used. The uncertainty
 in the energy of the tagged photons is mainly determined by the width
 of tagger focal-plane detectors and the energy of the MAMI electron beam
 used in the experiments. For the MAMI energies of 1508 and 1557~MeV,
 such an uncertainty was about $\pm 2$~MeV.

 The target was surrounded by a Particle IDentification
 (PID) detector~\cite{PID} used to distinguish between charged and
 neutral particles. It was made of 24 scintillator bars
 (50 cm long, 4 mm thick) arranged as a cylinder with a radius of 12 cm.

 The experimental trigger in the measurement with the 1508-MeV electron beam
 required the total energy deposited in the CB
 to exceed $\sim$320~MeV and the number of so-called hardware clusters
 in the CB (multiplicity trigger) to be larger than two.
 Depending on the data-taking period, events with cluster multiplicity two 
 were prescaled with a different rate.
 TAPS was not in the multiplicity trigger for these experiments.
 With the 1557-MeV electron beam, the trigger on the total energy
 in the CB was increased to $\sim$340~MeV.

 More details on the experimental conditions during the measurements
 with the 1508-MeV electron beam in 2007 are given
 in Refs.~\cite{slopemamic,etamamic}.

\section{Data handling}
\label{sec:Data}

 To search for a signal from $\eta \to \pi^0\gamma\gamma$ decay, 
 candidates for the process  $\gamma p\to \pi^0\gamma\gamma p\to 4\gamma p$
 were extracted from the analysis of events having
 five clusters (four from the photons and one from the proton)
 reconstructed in the CB and TAPS together.
 Four-cluster events, with only four photons detected,
 were neglected in the present analysis
 as the probability of non-detecting the final-state proton for the
 process $\gamma p\to \eta p\to \pi^0\gamma\gamma p$
 in these experiments was only about 20\%. Also, for such events,
 the proton information missing in the analysis resulted in
 a much stronger background contamination.

 The selection of event candidates was based on
 the kinematic-fit technique.
 The details of the kinematic-fit parameterization
 of the detector information and resolution are given
 in Ref.~\cite{slopemamic}. Events that satisfied
 the $\gamma p\to \pi^0\gamma\gamma p\to 4\gamma p$
 hypothesis with the confidence level (CL) greater than 5\%
 were accepted as possible reaction candidates.
 The kinematic-fit output was used to reconstruct the kinematics
 of the outgoing particles.

\begin{figure*}
\includegraphics[width=15.5cm,height=10.2cm,bbllx=1.cm,bblly=.0cm,bburx=19.5cm,bbury=13.3cm]{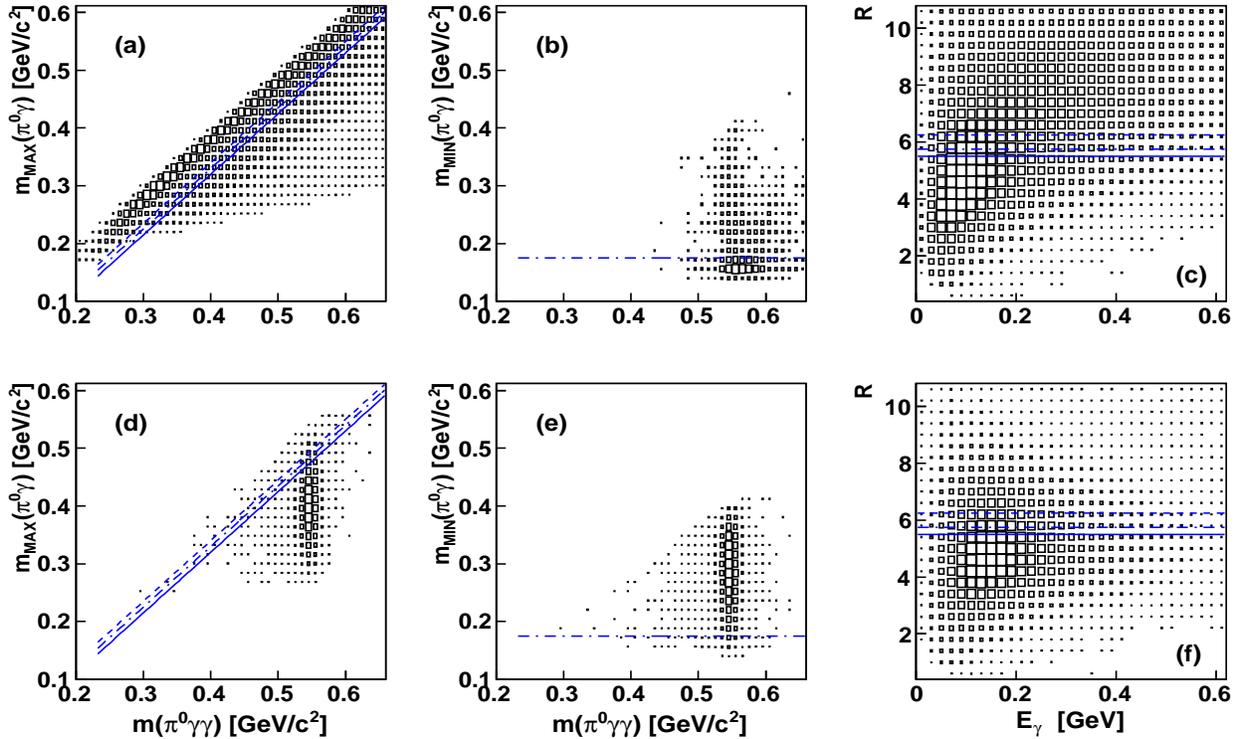}
\caption{ (Color online)
 Two-dimensional density distributions for events selected
 as the $\gamma p\to \pi^0\gamma\gamma p$ candidates
 obtained from the MC simulations of the three background
 processes (a) $\gamma p\to \pi^0\pi^0 p$,
 (b) $\gamma p\to \eta p \to \gamma\gamma p$,
 (c) $\gamma p\to \eta p \to 3\pi^0 p$ and (d), (e), (f) the process under the study
 $\gamma p\to \eta p \to \pi^0\gamma\gamma p$.
 Distributions (a), (d) plot the larger invariant mass $m(\pi^0\gamma)$
 and (b), (e) plot the smaller $m(\pi^0\gamma)$ 
 with respect to $m(\pi^0\gamma\gamma)$.
 Distributions (c), (f) plot the effective radii of the two clusters ascribed
 to the photons produced not from the $\pi^0$ decay.
 The cuts tested are shown by blue lines.
 Each of the three cuts on $m_{\mathrm{max}}(\pi^0\gamma)$ implies
 discarding all events that lie above the corresponding line.    
 The cuts on $m_{\mathrm{min}}(\pi^0\gamma)$ implies
 discarding all events that lie below the line.    
 Each of the three cuts on the cluster effective radius discards
 all events for which at least one
 of the two $R$ values is above the corresponding line.
}
 \label{fig:pi0gg_cuts} 
\end{figure*}

 The analysis technique and the selection criteria used
 in the present work were very similar to the ones
 described in detail in Refs.~\cite{CB_AGS,Prakhov_MENU07}.
 A signal from the $\eta \to \pi^0\gamma\gamma$ decay 
 was searched for as a peak in the invariant-mass
 spectrum of the $\pi^0\gamma\gamma$ final state, $m(\pi^0\gamma\gamma)$,
 at the mass region of the $\eta$ meson.
 Since the major contribution to the four-photon final state
 comes from the reaction $\gamma p\to \pi^0\pi^0 p$, it 
 was partially suppressed by eliminating all events for which
 the CL of satisfying the $\gamma p\to \pi^0\pi^0 p\to 4\gamma p$
 hypothesis was greater than 0.0001\%. Because of this cut,
 the measurement of the $\eta \to \pi^0\gamma\gamma$ decay is
 impossible for the region when the two-photon invariant mass
 is close to the $\pi^0$ mass. This cut would also eliminate
 $\eta \to \pi^0\pi^0$ decays if they exist. However, the
 $\eta \to \pi^0\pi^0$ decay is CP-violating and was never observed
 experimentally. Similar to the analysis of Ref.~\cite{CB_AGS},
 further suppression of the $\pi^0\pi^0$ background was achieved by
 cuts on the larger value of the two possible invariant masses $m(\pi^0\gamma)$
 with respect to $m(\pi^0\gamma\gamma)$.
 Three different cuts were tested. They are shown by three blue lines
 in Fig.~\ref{fig:pi0gg_cuts}(a), which plots the remaining background
 from the  Monte Carlo (MC) simulation for $\gamma p\to \pi^0\pi^0 p$,
 and in Fig.~\ref{fig:pi0gg_cuts}(d) depicting events
 from the MC simulation for $\gamma p\to \eta p \to \pi^0\gamma\gamma p$.
 Each of the three cuts on $m_{\mathrm{max}}(\pi^0\gamma)$ implies
 discarding all events that lie above the corresponding line.    
 Although the $\pi^0\pi^0$ background is comparably smooth
 in the region of the $\eta$ mass and cannot mimic a peak from
 $\eta \to \pi^0\gamma\gamma$ decays, the additional suppression
 of this background was important for observing
 a small $\eta \to \pi^0\gamma\gamma$ signal above statistical
 fluctuations in the background events.

 There are two background sources that can mimic a peak from
 $\eta \to \pi^0\gamma\gamma$ decays.
 These backgrounds are caused by $\eta \to \gamma\gamma$ 
 and $\eta \to 3\pi^0$ decays, which have much larger partial decay
 widths ($\Gamma(\eta\to\gamma\gamma)=39.31\%$ and
 $\Gamma(\eta\to3\pi^0)=32.57\%$~\cite{PDG_2012}) than
 $\eta \to \pi^0\gamma\gamma$. The $\eta \to \gamma\gamma$  decay
 can mimic the $\eta \to \pi^0\gamma\gamma$
 signal when the electromagnetic (e/m)
 showers of both outgoing photons split off.
 Similar to the analysis of Ref.~\cite{CB_AGS},
 this background was practically eliminated by a cut on the smaller value
 of the two possible invariant masses $m(\pi^0\gamma)$.
 This cut is shown by a blue line
 in Fig.~\ref{fig:pi0gg_cuts}(b), which plots the remaining background
 from the MC simulation for $\gamma p\to \eta p \to \gamma\gamma p$,
 and in Fig.~\ref{fig:pi0gg_cuts}(e) depicting events
 from the MC simulation for $\gamma p\to \eta p \to \pi^0\gamma\gamma p$.
 The cuts on $m_{\mathrm{min}}(\pi^0\gamma)$ implies
 discarding all events that lie below the line.    
 
 The $\eta \to 3\pi^0 \to 6\gamma$  decay mimics
 the $\eta \to \pi^0\gamma\gamma$ signal when the
 e/m showers of some outgoing photons overlap or are not detected.
 The case of overlapping e/m showers is more difficult to identify
 as the total energy of all the photons is conserved.
 As shown in Ref.~\cite{CB_AGS},
 the $\eta \to 3\pi^0$ background with overlapping e/m showers
 can be significantly suppressed by applying a cut on a so-called
 effective radius of the clusters, which are systematically wider
 in the case of overlapping showers. The effective radius $R$ of
 a cluster containing $k$ crystals with energy $E_i$
 deposited in crystal $i$ is defined as
 $R = \sqrt{\sum^k_i{E_i\cdot (\Delta r_i)^2}/\sum^k_i{E_i}}$,
 where $\Delta r_i$ is the opening angle (in degrees) between the cluster
 direction and the crystal axis.
 A density distribution of the cluster effective radii as a function
 of the cluster energy is shown in Fig.~\ref{fig:pi0gg_cuts}(c)
 for the MC simulation of $\gamma p\to \eta p \to 3\pi^0 p$
 and in Fig.~\ref{fig:pi0gg_cuts}(f)
 for the MC simulation of $\gamma p\to \eta p \to \pi^0\gamma\gamma p$.
 These distributions are plotted for the two clusters ascribed
 to the photons produced not from the $\pi^0$ decay.
 Three different cuts on the $R$ value were tested; they are shown by
 three blue lines in Figs.~\ref{fig:pi0gg_cuts}(c) and \ref{fig:pi0gg_cuts}(f).
 Each of the cuts discards all events for which at least one
 of the two $R$ values is above the corresponding line.

 Besides the so-called physical background, there are two more background sources.
 The first one comes from interactions of
 incident photons in the windows of the target cell.
 The subtraction of this background from
 experimental spectra is typically based on the 
 analysis of data samples that were taken
 with an empty (no liquid hydrogen) target.
 In the present analysis, the empty-target
 background was small and did not feature any visible
 $\eta$ peak in its $m(\pi^0\gamma\gamma)$ spectra. 
 Another background was caused by random coincidences
 of the tagger counts with the experimental trigger;
 its subtraction was carried out using 
 event samples for which all coincidences were random
 (see Refs.~\cite{slopemamic,etamamic} for more details).
\begin{figure*}
\includegraphics[width=15.5cm,height=5.5cm,bbllx=1.cm,bblly=.0cm,bburx=19.5cm,bbury=6.3cm]{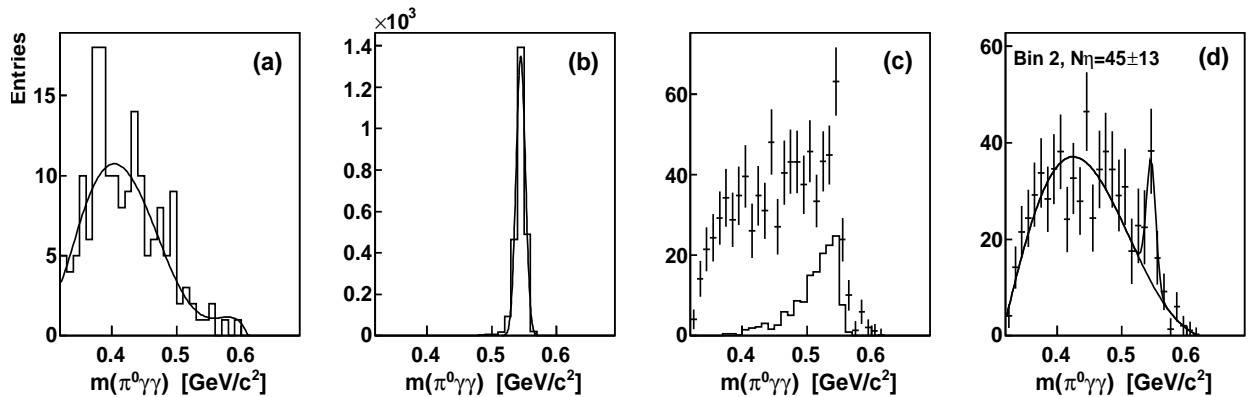}
\caption{
 $m(\pi^0\gamma\gamma)$ invariant-mass distributions obtained for 
 $m^2(\gamma\gamma)=(0.0375\pm0.0100)$~GeV$^2/c^4$:
 (a)~MC simulation of the background reaction $\gamma p\to \pi^0\pi^0p$
     with a polynomial fit;
 (b)~MC simulation of $\gamma p\to \eta p \to \pi^0\gamma\gamma p$
     with a Gaussian fit;
 (c)~experimental events from the 2009 data set (crosses) after subtracting
     the random background; the $\gamma p\to \eta p \to 3\pi^0 p$ background
     (solid line) expected to remain after all cuts (solid line);
 (d)~experimental events from (c) after subtracting
     the $\gamma p\to \eta p \to 3\pi^0 p$ background
     fitted with the sum of a Gaussian and a polynomial.
}
 \label{fig:pi0ggz5_fit_bin2_4x1} 
\end{figure*}
\begin{figure*}
\includegraphics[width=15.5cm,height=5.5cm,bbllx=1.cm,bblly=.0cm,bburx=19.5cm,bbury=6.3cm]{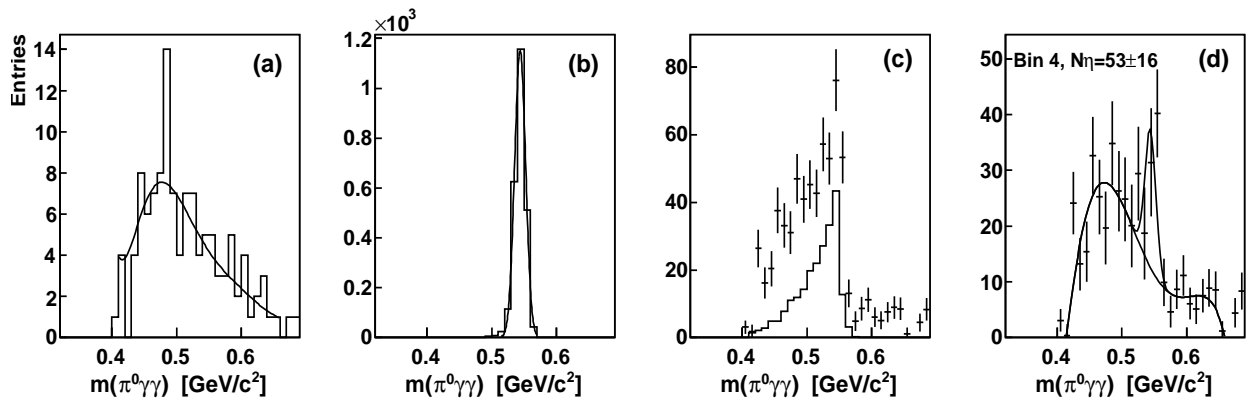}
\caption{
 Same as Fig.~\protect\ref{fig:pi0ggz5_fit_bin2_4x1}, but
 for $m^2(\gamma\gamma)=(0.0775\pm0.0100)$~GeV$^2/c^4$.
}
 \label{fig:pi0ggz5_fit_bin4_4x1} 
\end{figure*}
%

 The selection criteria were optimized using MC
 simulations of the process $\gamma p\to \eta p \to \pi^0\gamma\gamma p$
 and all the background reactions.  
 To reproduce the experimental yield of the $\eta$ mesons as a function
 of the incident-photon energy, the $\gamma p\to \eta p$ reaction
 was generated according to its excitation function,
 measured in the same experiment~\cite{etamamic}, which was then
 folded with the bremsstrahlung energy dependence of the incident photons. 
 Since the photon-beam energy range used in the analysis is large
 (nearly 700~MeV), the $\gamma p\to \eta p$ production angular distribution
 changes with energy. However, the production angle was generated
 isotropically as its experimental distribution, averaged over all energies,
 was sufficiently close to isotropic.
 The $\eta \to \pi^0\gamma\gamma$ decay was generated according
 to the matrix element of the transition amplitude from Ref.~\cite{Ng1},
 calculated assuming vector-meson dominance.
 The $d\Gamma(\eta\to\pi^0\gamma\gamma)/dm^2(\gamma\gamma)$ dependence from
 this amplitude was found to be close to the experimental
 data~\cite{CB_AGS,Prakhov_MENU07}.
 The MC simulation of the $\gamma p\to \pi^0\pi^0p$ reaction was done
 in the same way as reported in Ref.~\cite{p2pi0mamic}.
 
 For all reactions, the generated events
 were propagated through a {\sc GEANT} (version 3.21) simulation of the experimental
 setup. To reproduce resolutions of the experimental data,
 the {\sc GEANT} output (energy and timing) was subject
 to additional smearing, allowing both the simulated and experimental data
 to be analyzed in the same way.
 The simulated events were also tested for whether they passed the trigger
 requirements. From the analysis of the MC simulation, the average acceptance
 for the process $\gamma p\to \eta p \to \pi^0\gamma\gamma p$
 varied between 7\% and 14\%, depending on the criteria used for event selection.
\begin{figure*}
\includegraphics[width=15.5cm,height=5.5cm,bbllx=1.cm,bblly=.0cm,bburx=19.5cm,bbury=6.3cm]{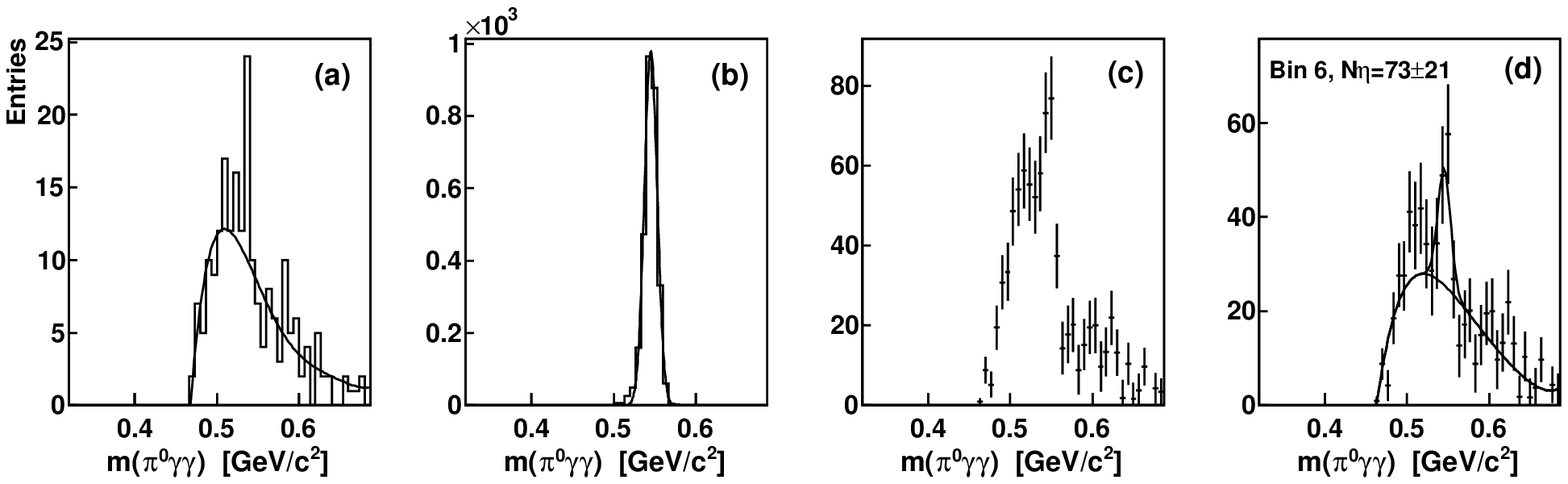}
\caption{
 Same as Fig.~\protect\ref{fig:pi0ggz5_fit_bin2_4x1}, but
 for $m^2(\gamma\gamma)=(0.11875\pm0.01125)$~GeV$^2/c^4$.
}
 \label{fig:pi0ggz5_fit_bin6_4x1} 
\end{figure*}

 To measure the $\eta \to \pi^0\gamma\gamma$ decay rate as a function
 of the diphoton invariant mass squared, the data were
 divided into eight $m^2(\gamma\gamma)$ bins.
 The bin corresponding to the the $\pi^0$ mass was not analyzed since
 it was almost empty due to suppressing $\gamma p\to \pi^0\pi^0 p\to 4\gamma p$
 events by the 0.0001\% cut on the kinematic-fit CL for this reaction. 
 The number of $\eta \to \pi^0\gamma\gamma$ decays observed
 in each $m^2(\gamma\gamma)$ bin was determined by fitting
 an $\eta$ peak above a relatively smooth background, containing
 mostly the $\pi^0\pi^0$ events survived the cuts.
 The fitting procedure is illustrated in Fig.~\ref{fig:pi0ggz5_fit_bin2_4x1}
 for bin $m^2(\gamma\gamma)=(0.0375\pm0.0100)$~GeV$^2/c^4$ of the 2009 data set.
 Figure~\ref{fig:pi0ggz5_fit_bin2_4x1}(a) depicts the $m(\pi^0\gamma\gamma)$
 invariant-mass distribution for the MC simulation of
 the background reaction $\gamma p\to \pi^0\pi^0p$ fitted with a polynomial.
 Figure~\ref{fig:pi0ggz5_fit_bin2_4x1}(b) shows the same distribution
 for the MC simulation of $\gamma p\to \eta p \to \pi^0\gamma\gamma p$
 fitted with a Gaussian.
 The experimental distribution after subtracting the random
 background is shown by crosses in Fig.~\ref{fig:pi0ggz5_fit_bin2_4x1}(c).
 The MC simulation for the $\gamma p\to \eta p \to 3\pi^0 p$ background 
 is shown in the same figure by a solid line. The normalization of this
 background is based on the ratio of the number of
 $\gamma p\to \eta p \to 3\pi^0 p$ events produced in this experiment
 to the number of events generated for this process in the MC simulation.
 The experimental distribution after subtraction of
 the $\gamma p\to \eta p \to 3\pi^0 p$ background is
 shown in Fig.~\ref{fig:pi0ggz5_fit_bin2_4x1}(d).
 This distribution was fitted with the sum of a Gaussian
 and a polynomial. The mean value and $\sigma$ of the Gaussian
 were fixed to the values obtained from the previous fit
 to the MC simulation for $\gamma p\to \eta p \to \pi^0\gamma\gamma p$.
 The initial values for polynomial coefficients were taken equal to
 the output parameters of the polynomial fit to the $\gamma p\to \pi^0\pi^0p$
 MC simulation. The order of the polynomial was chosen to be sufficient for
 fairly good description of the background distribution in its full
 $m(\pi^0\gamma\gamma)$ range. Typically, a polynomial of
 order seven was sufficient for fitting to background spectra
 of lower $m^2(\gamma\gamma)$ bins, having wider $m(\pi^0\gamma\gamma)$
 distributions. While, for the bins with the highest $m^2(\gamma\gamma)$ values,
 having quite narrow $m(\pi^0\gamma\gamma)$ distributions, the fit with
 a polynomial of order four provided a sufficient quality.
 
 The experimental number of $\eta \to \pi^0\gamma\gamma$ decays in
 the $m(\pi^0\gamma\gamma)$ distribution shown
 in Fig.~\ref{fig:pi0ggz5_fit_bin2_4x1}(d) was determined from
 the area under the Gaussian. For consistency,  
 the corresponding detection efficiency was calculated
 in the same way (i.e, based on a Gaussian fit to the MC simulation
 for $\gamma p\to \eta p \to \pi^0\gamma\gamma p$, instead of just using
 the number of entries in the $m(\pi^0\gamma\gamma)$ distribution).
 The uncertainty in the number of $\eta \to \pi^0\gamma\gamma$ decays
 observed, which is calculated from the fit results, does not
 reflect the actual statistic in the signal peak.
 This uncertainty is much larger because of the background remaining
 under the $\eta \to \pi^0\gamma\gamma$ peak and the increase in
 the error bars in the experimental distribution after subtracting
 the random and the $\eta \to 3\pi^0$ backgrounds. 
\begin{table*}
\caption
[tab:etatff]{
 $d\Gamma(\eta \to \pi^0\gamma\gamma)/dm^2(\gamma\gamma)$ results of this work
 obtained from the data of 2007, 2009, and their average
 } \label{tab:results}
\begin{ruledtabular}
\begin{tabular}{|c|c|c|c|c|} 
\hline
 $m^2(\gamma\gamma)$~[GeV$^2/c^4$]
 & $0.-0.011$ & $0.0275-0.0475$ & $0.0475-0.0675$ & $0.0675-0.0875$ \\
\hline
 2007 
 & $3.69\pm1.60_{\mathrm{stat}}\pm0.44_{\mathrm{syst}}$
 & $1.88\pm0.76_{\mathrm{stat}}\pm0.27_{\mathrm{syst}}$
 & $2.33\pm0.64_{\mathrm{stat}}\pm0.28_{\mathrm{syst}}$
 & $2.02\pm0.54_{\mathrm{stat}}\pm0.23_{\mathrm{syst}}$ \\
 & $3.69\pm1.66_{\mathrm{tot}}$ & $1.88\pm0.81_{\mathrm{tot}}$
 & $2.33\pm0.70_{\mathrm{tot}}$ & $2.02\pm0.59_{\mathrm{tot}}$ \\
\hline
 2009 
 & $3.67\pm1.51_{\mathrm{stat}}\pm0.62_{\mathrm{syst}}$
 & $2.15\pm0.74_{\mathrm{stat}}\pm0.29_{\mathrm{syst}}$
 & $2.12\pm0.65_{\mathrm{stat}}\pm0.15_{\mathrm{syst}}$
 & $2.01\pm0.64_{\mathrm{stat}}\pm0.17_{\mathrm{syst}}$ \\
 & $3.67\pm1.63_{\mathrm{tot}}$ & $2.15\pm0.80_{\mathrm{tot}}$
 & $2.12\pm0.66_{\mathrm{tot}}$ & $2.01\pm0.66_{\mathrm{tot}}$ \\
\hline
 2007+2009 
 & $3.68\pm1.16_{\mathrm{tot}}$ & $2.02\pm0.57_{\mathrm{tot}}$
 & $2.22\pm0.48_{\mathrm{tot}}$ & $2.01\pm0.44_{\mathrm{tot}}$ \\
\hline
\hline
 $m^2(\gamma\gamma)$~[GeV$^2/c^4$]
 & $0.0875-0.1075$ & $0.1075-0.13$ & $0.13-0.17$ & \\
\hline
 2007 
 & $1.87\pm0.47_{\mathrm{stat}}\pm0.23_{\mathrm{syst}}$
 & $1.70\pm0.48_{\mathrm{stat}}\pm0.12_{\mathrm{syst}}$
 & $0.89\pm0.26_{\mathrm{stat}}\pm0.13_{\mathrm{syst}}$
 & \\
 & $1.87\pm0.53_{\mathrm{tot}}$ & $1.70\pm0.49_{\mathrm{tot}}$ & $0.89\pm0.29_{\mathrm{tot}}$ & \\
\hline
 2009 
 & $2.12\pm0.63_{\mathrm{stat}}\pm0.15_{\mathrm{syst}}$
 & $1.86\pm0.52_{\mathrm{stat}}\pm0.16_{\mathrm{syst}}$
 & $1.08\pm0.30_{\mathrm{stat}}\pm0.25_{\mathrm{syst}}$
 & \\
 & $2.12\pm0.65_{\mathrm{tot}}$ & $1.86\pm0.54_{\mathrm{tot}}$ & $1.08\pm0.39_{\mathrm{tot}}$ & \\
\hline
 2007+2009 
 & $1.97\pm0.41_{\mathrm{tot}}$ & $1.77\pm0.36_{\mathrm{tot}}$ & $0.95\pm0.23_{\mathrm{tot}}$ & \\
\hline
\end{tabular}
\end{ruledtabular}
\end{table*}

 Figures~\ref{fig:pi0ggz5_fit_bin4_4x1} and \ref{fig:pi0ggz5_fit_bin6_4x1}
 illustrate the fitting procedure for two more bins of higher $m^2(\gamma\gamma)$ masses,
 showing changes in the shape of the $\pi^0\pi^0$
 and the $\eta \to 3\pi^0$ background contaminations under
 the $\eta \to \pi^0\gamma\gamma$ signal. 

\section{Experimental results}
  \label{sec:Results}

 For every fit to the experimental $m(\pi^0\gamma\gamma)$ distributions
 (see Section~\ref{sec:Data} for details), the number of
 $\eta \to \pi^0\gamma\gamma$ decays initially produced
 was obtained by dividing the number of $\eta \to \pi^0\gamma\gamma$ decays
 determined from the fit by the corresponding detection efficiency.   
 Values of $d\Gamma(\eta\to\pi^0\gamma\gamma)/dm^2(\gamma\gamma)$
 for every fit were obtained from the initial number of $\eta \to \pi^0\gamma\gamma$
 by taking into account the full decay width
 $\Gamma_{\eta}=(1.30\pm0.07)$~keV~\cite{PDG_2012}, the total number of $\eta$ mesons
 produced in the experiment, and the width of the corresponding $m^2(\gamma\gamma)$ bin.
 The uncertainty in an individual $d\Gamma(\eta\to\pi^0\gamma\gamma)/dm^2(\gamma\gamma)$
 value was obtained in the same way from the uncertainty in
 the number of $\eta \to \pi^0\gamma\gamma$ decays determined from every fit.
 The calculation of the total number of $\eta$ mesons produced in the experiment
 was based on the analysis of the process $\gamma p\to \eta p \to 3\pi^0p$
 in both the 2007~\cite{etamamic} and 2009 data sets, using $0.3257\pm 0.0023$
 for the $\eta \to 3\pi^0$ branching ratio~\cite{PDG_2012}. 
 Since the data of 2007 and 2009 were taken under different experimental
 conditions, both the data sets were analyzed independently.  
 As a signal from $\eta \to \pi^0\gamma\gamma$ decays in each $m^2(\gamma\gamma)$ bin
 was quite small, compared to the magnitude of the error bars in the final
 $m(\pi^0\gamma\gamma)$ spectra, the fitting procedure was repeated many times
 for each $m^2(\gamma\gamma)$ bin, testing different combinations of selection criteria.
 This provided a check on the stability of
 the $d\Gamma(\eta\to\pi^0\gamma\gamma)/dm^2(\gamma\gamma)$ results,
 the average of which was used to obtain a more reliable
 $d\Gamma(\eta\to\pi^0\gamma\gamma)/dm^2(\gamma\gamma)$ value
 for each $m^2(\gamma\gamma)$ bin.

 Three different cuts on $m_{\mathrm{max}}(\pi^0\gamma)$, shown
 in Figs.~\ref{fig:pi0gg_cuts}(a) and \ref{fig:pi0gg_cuts}(d),
 were used to change the level of the $\pi^0\pi^0$ background
 under the $\eta \to \pi^0\gamma\gamma$ signal in the bins
 with lower $m^2(\gamma\gamma)$ masses.
 Since the production of $\gamma p\to \eta p$~\cite{etamamic}
 is mostly accumulated in the region of $S_{11}(1535)$, whereas
 the $\gamma p\to \pi^0\pi^0p$ total cross section changes much less between
 the $\eta$ threshold ($E_\gamma=0.707$~GeV) and the incident-photon
 energy $E_\gamma=1.4$~GeV~\cite{p2pi0mamic}, different cuts on $E_\gamma$
 were used to change the ratio of the $\eta \to \pi^0\gamma\gamma$ signal
 to the $\pi^0\pi^0$ background in each $m^2(\gamma\gamma)$ bin.
 Three different cuts on the cluster effective radius, shown
 in Figs.~\ref{fig:pi0gg_cuts}(c) and \ref{fig:pi0gg_cuts}(f),
 were used to change the ratio of the $\eta \to \pi^0\gamma\gamma$ signal
 to the $\eta \to 3\pi^0$ background.
 Different cuts on the kinematic-fit CL (such as 5\%, 10\%, 15\%, and 20\%)
 for the $\gamma p\to \pi^0\gamma\gamma p\to 4\gamma p$ hypothesis were used
 to change the ratio of the $\eta \to \pi^0\gamma\gamma$ signal to
 both the $\eta \to 3\pi^0$ and $\pi^0\pi^0$ background.

 Since the same data were used to obtain 
 the $d\Gamma(\eta \to \pi^0\gamma\gamma)/dm^2(\gamma\gamma)$ values
 after applying different selection criteria, all results
 corresponding to an individual $m^2(\gamma\gamma)$ bin  were correlated.
 Although all those results were in agreement within their uncertainties,
 the magnitude of the uncertainties themselves did not allow
 a careful study of the systematics in the results.  
 In the end, the calculation of the final
 $d\Gamma(\eta \to \pi^0\gamma\gamma)/dm^2(\gamma\gamma)$ values
 and their uncertainties was carried out in the same way as
 in the analyses of Refs.~\cite{CB_AGS,Prakhov_MENU07}.
 The $d\Gamma(\eta \to \pi^0\gamma\gamma)/dm^2(\gamma\gamma)$ value
 in one $m^2(\gamma\gamma)$ bin was obtained by averaging the results
 of all tests made for this bin. Then, the average of the corresponding
 $d\Gamma(\eta \to \pi^0\gamma\gamma)/dm^2(\gamma\gamma)$ uncertainties,
 calculated from the fit errors, was considered, for simplicity,
 as the statistical uncertainty of
 the $d\Gamma(\eta \to \pi^0\gamma\gamma)/dm^2(\gamma\gamma)$ value
 in this bin. Whereas, the systematic uncertainty of the
 $d\Gamma(\eta \to \pi^0\gamma\gamma)/dm^2(\gamma\gamma)$ value
 was taken as the root mean square of the results from all tests
 made for this bin. The total uncertainty of
 the $d\Gamma(\eta \to \pi^0\gamma\gamma)/dm^2(\gamma\gamma)$ value
 in each $m^2(\gamma\gamma)$ bin
 was calculated by adding in quadrature its statistical and
 systematic uncertainties.
 In the end, the $d\Gamma(\eta \to \pi^0\gamma\gamma)/dm^2(\gamma\gamma)$
 results from 2007 and 2009, which were independent measurements,
 were combined as a weighted average with weights taken as inverse values
 of their total uncertainties in quadrature.
 The numerical $d\Gamma(\eta \to \pi^0\gamma\gamma)/dm^2(\gamma\gamma)$ values
 obtained from the data of 2007, 2009, and their average are listed 
 in Table~\ref{tab:results}.

\section{Discussion of the results}
  \label{sec:Discussion}

\begin{figure}
\includegraphics[width=6.5cm,height=7.cm,bbllx=2.cm,bblly=.5cm,bburx=18.cm,bbury=17.5cm]{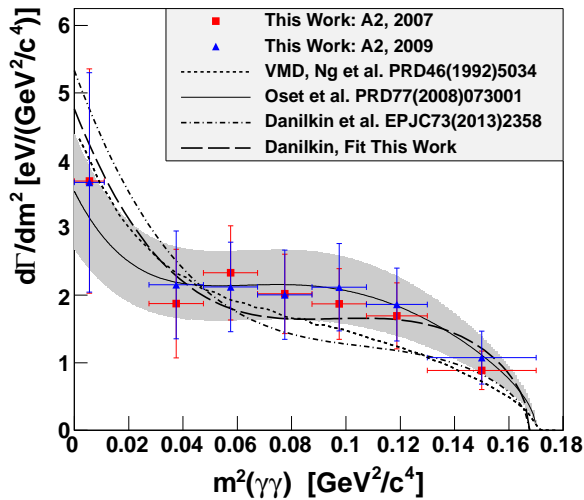}
\caption{(Color online)
  Comparison of the $d\Gamma(\eta \to \pi^0\gamma\gamma)/dm^2(\gamma\gamma)$
 results obtained from the analyses of the 2007 (red squares) data set
 with the results from 2009 (blue triangles)
 and with the calculations of Ref.~\protect\cite{Ng1} (short-dashed line),
 Ref.~\protect\cite{Oset_2008} (solid line with shaded error band),
 Ref.~\protect\cite{Ch_Lag_2012} (dash-dotted line), and
 Ref.~\protect\cite{Ch_Lag_refit} (long-dashed line) fitting to the present data only.
}
 \label{fig:dGdm2_etapi0gg_A2_2007_2009_th} 
\end{figure}

\begin{figure}
\includegraphics[width=6.5cm,height=7.cm,bbllx=2.cm,bblly=.5cm,bburx=18.cm,bbury=17.5cm]{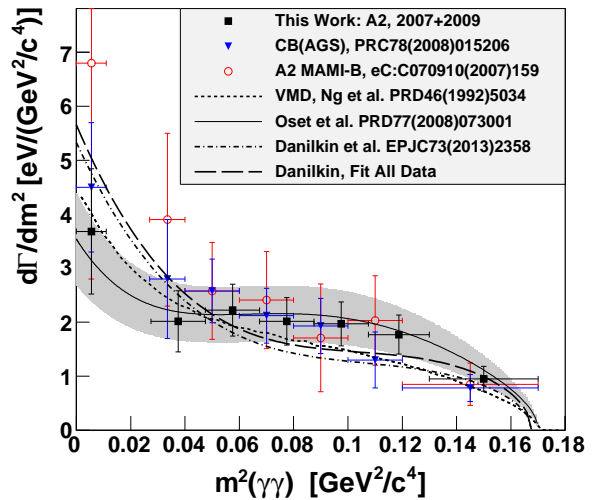}
\caption{(Color online)
  Comparison of the combined $d\Gamma(\eta \to \pi^0\gamma\gamma)/dm^2(\gamma\gamma)$
 results from the present analysis (black squares) with previous measurements by
 the CB at AGS~\protect\cite{CB_AGS} (blue triangles) and 
 the A2 at MAMI-B~\protect\cite{Prakhov_MENU07} (red circles),
 and with the calculations of Ref.~\protect\cite{Ng1} (short-dashed line),
 Ref.~\protect\cite{Oset_2008} (solid line with shaded error band),
 Ref.~\protect\cite{Ch_Lag_2012} (dash-dotted line), and
 Ref.~\protect\cite{Ch_Lag_refit} (long-dashed line) fitting to the present data along with
 the data points of Refs.~\protect\cite{CB_AGS,Prakhov_MENU07}.
}
 \label{fig:dGdm2_etapi0gg_A2_exp_th} 
\end{figure}

 The individual $d\Gamma(\eta \to \pi^0\gamma\gamma)/dm^2(\gamma\gamma)$ results
 from the two measurements in 2007 and 2009
 are plotted in Fig.~\ref{fig:dGdm2_etapi0gg_A2_2007_2009_th},
 demonstrating very good agreement of the two data sets within the error bars that
 correspond to the total uncertainties.
 In Fig.~\ref{fig:dGdm2_etapi0gg_A2_exp_th}, the results combined from
 2007 and 2009 are compared to the existing data~\cite{CB_AGS,Prakhov_MENU07},
 again, within the error bars, showing good agreement with the previous measurements.
 As also seen, the uncertainties of the combined results from the present work
 are smaller than the uncertainties of the results from
 Refs.~\cite{CB_AGS,Prakhov_MENU07}.

 In Figs.~\ref{fig:dGdm2_etapi0gg_A2_2007_2009_th} and \ref{fig:dGdm2_etapi0gg_A2_exp_th},
 the data points are also compared to calculations from different models.
 The VMD calculation from Ref.~\cite{Ng1} is shown by a short-dashed line;
 the matrix element of the corresponding transition amplitude was used to generate
 $\eta \to \pi^0\gamma\gamma$ decays in the MC simulation made
 for the acceptance calculation.
 The revised calculation based on a chiral unitary approach~\cite{Oset_2008}
 is shown by a solid line with a shaded error band.
 The calculation involving a theoretical study of photon-fusion reactions
 based on a chiral Lagrangian with dynamical light vector mesons~\cite{Ch_Lag_2012}
 is shown by a dash-dotted line.

 The impact of the present results
 on models was tested by repeating the analysis of photon-fusion
 reactions after including the new
 $d\Gamma(\eta\to\pi^0\gamma\gamma)/dm^2(\gamma\gamma)$ data
 in their fits~\cite{Ch_Lag_refit}. Their first fit,
 in which only the new data from this work were used,
 is shown in Fig.~\ref{fig:dGdm2_etapi0gg_A2_2007_2009_th} by a long-dashed line.
 The second fit, shown by a long-dashed line in Fig.~\ref{fig:dGdm2_etapi0gg_A2_exp_th},
 involves the present data together with the data points from
 Refs.~\cite{CB_AGS,Prakhov_MENU07}.
 As seen, the results of the second fit are much closer to the original
 fit~\cite{Ch_Lag_2012} than the results of the first fit.
 These two fits resulted in a new determination of the two low-energy
 constants, $g_3$ and $h_O$~\footnotemark, of the chiral vector-meson
 Lagrangian~\cite{Ch_Lag_2012}, in which these constants parametrize
 contact interactions between two pseudoscalar and two vector mesons.
\footnotetext{The first fit yields values $g_3=-3.69$ and $h_O=4.10$
 for the Lagrangian parameters, compared to $g_3=-4.62$ and $h_O=3.71$
 from the second fit and to $g_3=-4.88$ and $h_O=3.27$
 from Ref.~\protect\cite{Ch_Lag_2012}.}
 Despite the impact of the present data on the parameters of the chiral
 Lagrangian, the results for the photon-fusion reactions remained
 quite stable with respect to these changes and so do not affect 
 any conclusions made in Ref.~\cite{Ch_Lag_2012}.

 As seen in Fig.~\ref{fig:dGdm2_etapi0gg_A2_2007_2009_th}, all the calculations
 go through the error bars of the individual 2007 and 2009 data points.
 The combined results from the 2007 and 2009 data sets, shown in
 Fig.~\ref{fig:dGdm2_etapi0gg_A2_exp_th} , have smaller
 uncertainties and, in our opinion, overlap better with the revised calculation
 based on a chiral unitary approach~\cite{Oset_2008}.
 Nevertheless, the magnitude of the experimental uncertainties is still not
 sufficient to rule out any of the theoretical calculations shown,
 and the need for more accurate measurements is obvious. 

 In all previous experiments, the partial decay width
 $\Gamma(\eta \to \pi^0\gamma\gamma)$ was determined via measuring
 the total number of $\eta \to \pi^0\gamma\gamma$ decays observed
 and correcting it with the detection efficiency.
 However, if the $d\Gamma(\eta \to \pi^0\gamma\gamma)/dm^2(\gamma\gamma)$
 dependence has been already measured, the partial decay width
 can be determined just by integrating such a distribution,
 where the value and its uncertainty for the missing bin at the $\pi^0$ mass
 can be extrapolated from the values and uncertainties of the adjacent bins.
 If, for simplicity, a linear extrapolation is used for the missing bin,
 integrating the experimental
 $d\Gamma(\eta \to \pi^0\gamma\gamma)/dm^2(\gamma\gamma)$ distribution
 from this work results in
 $$\Gamma(\eta \to \pi^0\gamma\gamma)=(0.330\pm0.030_{\mathrm{tot}})~\mathrm{eV},$$
 where the total uncertainty is calculated by using the error-propagation
 formula for a sum of values with independent errors.
\begin{figure*}
\includegraphics[width=15.5cm,height=5.5cm,bbllx=1.cm,bblly=.0cm,bburx=19.5cm,bbury=6.3cm]{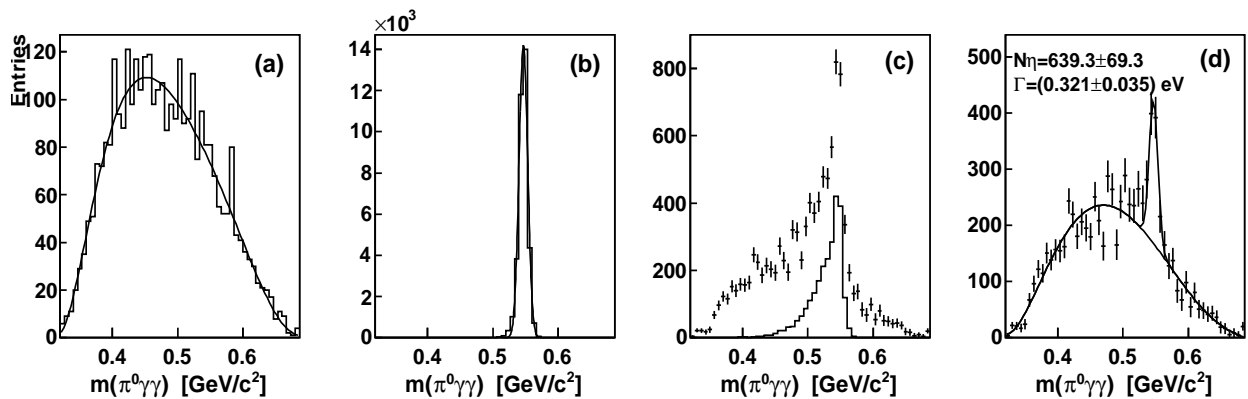}
\caption{
 Same as Fig.~\protect\ref{fig:pi0ggz5_fit_bin2_4x1}, but
 for the full range of $m^2(\gamma\gamma)$ from the 2007 data set
 and using a looser cut on the cluster effective radius,
 shown by the dashed blue line in
 Figs.~\protect\ref{fig:pi0gg_cuts}(c) and \protect\ref{fig:pi0gg_cuts}(f). 
}
 \label{fig:pi0ggz5_fit_rcut1_4x1} 
\end{figure*}
\begin{figure*}
\includegraphics[width=15.5cm,height=5.5cm,bbllx=1.cm,bblly=.0cm,bburx=19.5cm,bbury=6.3cm]{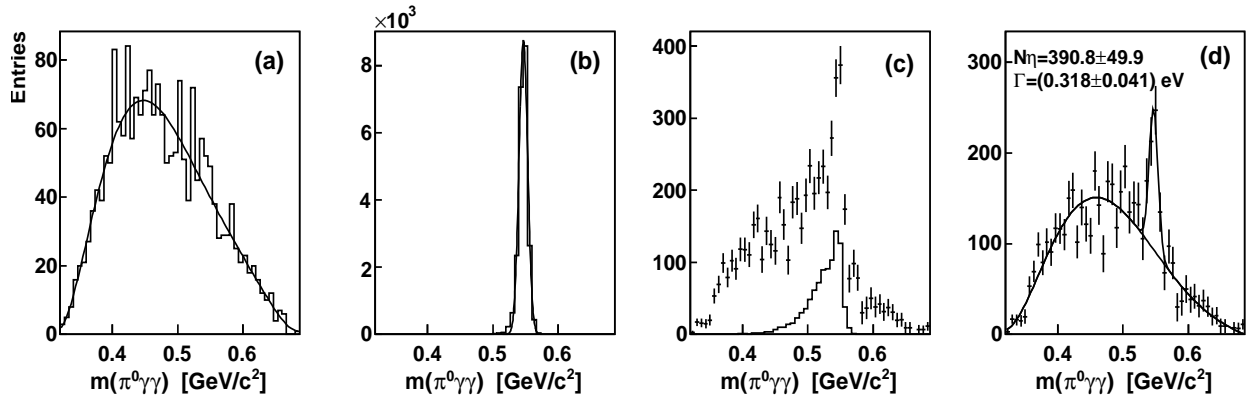}
\caption{
 Same as Fig.~\protect\ref{fig:pi0ggz5_fit_rcut1_4x1}, but using
 a tighter cut on the cluster effective radius,
 shown by the solid blue line in
 Figs.~\protect\ref{fig:pi0gg_cuts}(c) and \protect\ref{fig:pi0gg_cuts}(f).
}
 \label{fig:pi0ggz5_fit_rcut3_4x1} 
\end{figure*}

 The determination of $\Gamma(\eta \to \pi^0\gamma\gamma)$
 in the so-called traditional approach would again require repeating the fitting
 procedure for the full $m^2(\gamma\gamma)$ range with different combinations
 of selection criteria to evaluate the average result and its statistical and
 systematic uncertainties and to combine the results from the 2007 and 2009
 data sets in the end.
 To demonstrate that the traditional approach gives similar results,
 the corresponding fitting procedure is illustrated in
 Figs.~\ref{fig:pi0ggz5_fit_rcut1_4x1} and \ref{fig:pi0ggz5_fit_rcut3_4x1}
 for the 2007 data set with two different cuts on the cluster effective radius.
 Figure~\ref{fig:pi0ggz5_fit_rcut1_4x1} corresponds to a looser effective-radius cut
 (shown by the dashed blue line in
 Figs.~\ref{fig:pi0gg_cuts}(c) and \ref{fig:pi0gg_cuts}(f)),
 which provides a poorer ratio of the signal to the $\eta \to 3\pi^0$ background,
 but leaves a greater statistic for the $\eta \to \pi^0\gamma\gamma$ decays.
 The number of $\eta \to \pi^0\gamma\gamma$ decays found for this case
 is $639\pm59$, and the partial decay width is calculated to be
 $\Gamma(\eta \to \pi^0\gamma\gamma)=(0.321\pm0.035_{\mathrm{stat}})$~eV.
 Figure~\ref{fig:pi0ggz5_fit_rcut3_4x1} corresponds to a tighter effective-radius cut
 (shown by the solid blue line in
 Figs.~\ref{fig:pi0gg_cuts}(c) and \ref{fig:pi0gg_cuts}(f)),
 which provides a better ratio of the signal to the $\eta \to 3\pi^0$ background,
 but leaves a lower statistic for the $\eta \to \pi^0\gamma\gamma$ decays.
 The number of $\eta \to \pi^0\gamma\gamma$ decays found here
 is only $391\pm50$, and the decay width,
 $\Gamma(\eta \to \pi^0\gamma\gamma)=(0.318\pm0.041_{\mathrm{stat}})$~eV,
 has almost the same value but with a larger uncertainty,
 hinting on the smallness of the systematics because of applying
 cuts on a value for $R$.
 As also seen, the results of the traditional approach for determining
 $\Gamma(\eta \to \pi^0\gamma\gamma)$ are in good agreement with the value
 obtained from integrating the 
 $d\Gamma(\eta \to \pi^0\gamma\gamma)/dm^2(\gamma\gamma)$ data points. 
  
 The result obtained in this work for the partial decay width,
 $\Gamma(\eta \to \pi^0\gamma\gamma)=(0.330\pm0.030_{\mathrm{tot}})$~eV,
 is in good agreement with the present PDG~\cite{PDG_2012} value,
 $\Gamma(\eta\to\pi^0\gamma\gamma)=(0.35\pm0.07)$~eV, 
 and with the values from the calculations
 shown in Figs.~\ref{fig:dGdm2_etapi0gg_A2_2007_2009_th}
 and \ref{fig:dGdm2_etapi0gg_A2_exp_th}.
 The calculation based on the VMD transition amplitude
 from Ref.~\cite{Ng1} gives $\Gamma(\eta\to\pi^0\gamma\gamma)=0.30^{+0.16}_{-0.13}$~eV.
 The calculation based on a chiral unitary approach~\cite{Oset_2008}
 results in $\Gamma(\eta \to \pi^0\gamma\gamma)=(0.33\pm0.08)$~eV.
 The results from Refs.~\cite{Ch_Lag_2012,Ch_Lag_refit}
 are not considered as their unknown parameters were fitted
 to existing $d\Gamma(\eta \to \pi^0\gamma\gamma)/dm^2(\gamma\gamma)$ data.

 Note that all the calculations shown in
 Figs.~\ref{fig:dGdm2_etapi0gg_A2_2007_2009_th}
 and \ref{fig:dGdm2_etapi0gg_A2_exp_th} give very similar results for
 $\Gamma(\eta\to\pi^0\gamma\gamma)$, but their
 $d\Gamma(\eta \to \pi^0\gamma\gamma)/dm^2(\gamma\gamma)$ distributions
 are quite different. This demonstrates that a precision measurement of
 $d\Gamma(\eta \to \pi^0\gamma\gamma)/dm^2(\gamma\gamma)$ would be
 a more efficient way to test the reliability of $\chi$PTh calculations
 than measuring just the $\Gamma(\eta\to\pi^0\gamma\gamma)$ value. 

\section{Summary and conclusions}
\label{sec:Conclusion}

 A new measurement of the rare, doubly radiative decay
 $\eta\to\pi^0\gamma\gamma$ was conducted by the A2
 collaboration at MAMI. The results are based
 on analysis of $1.2\times 10^3 ~\eta \to \pi^0\gamma\gamma$
 decays from a total of $6 \times 10^7$ $\eta$ mesons produced in
 the $\gamma p\to \eta p$ reaction.
 The statistical accuracy of the new results for
 the $d\Gamma(\eta \to \pi^0\gamma\gamma)/dm^2(\gamma\gamma)$
 dependence and the partial decay width,
 $\Gamma(\eta \to \pi^0\gamma\gamma) = (0.33\pm 0.03_{\mathrm{tot}})$~eV,
 is better than all previous measurements of $\eta\to\pi^0\gamma\gamma$.
 The present results for $d\Gamma(\eta\to\pi^0\gamma\gamma)/dm^2(\gamma\gamma)$
 are in good agreement with previous measurements and recent theoretical calculations
 for this dependence. 
 
\section*{Acknowledgments}
 This paper is devoted to the memory of our appreciated colleague
 B.M.K.~Nefkens who passed away just before the paper submission.
 B.M.K.~Nefkens was the driving force behind the research program
 dedicated to $\eta$-meson decays with the Crystal Ball at MAMI.

 The authors wish to acknowledge the excellent support of the accelerator
 group and operators of MAMI.
 This work was supported by the Deutsche Forschungsgemeinschaft (SFB443,
 SFB/TR16, and SFB1044), DFG-RFBR (Grant No. 09-02-91330), the European Community-Research
Infrastructure Activity under the FP6 ``Structuring the European Research Area''
program (Hadron Physics, Contract No. RII3-CT-2004-506078), Schweizerischer
Nationalfonds, the UK Science and Technology Facilities Council
(Grant No. STFC 57071/1, 50727/1), the U.S. Department of Energy
 and National Science Foundation,  INFN (Italy), and NSERC (Canada). 
The work of I.V.~Danilkin is supported by the U.S. Department of Energy
(Contract No. DE-AC05-06OR23177).
A.~Fix acknowledges additional support from the Russian Federation federal program
``Kadry'' (Contract No. P691) and the MSE Program ``Nauka'' (Contract No. 1.604.2011).
We thank the undergraduate students of Mount Allison University
and The George Washington University for their assistance.

\end{document}